%% file: main.tex
\documentclass[sigconf]{acmart}

\usepackage{todonotes}
\usepackage{xspace}
\usepackage{colortbl}
\usepackage{hyperref}
\usepackage{url}
\usepackage{multirow}
\usepackage{balance}

\usepackage[T1]{fontenc}

\usepackage{pifont}

\usepackage[para]{footmisc}
\usepackage{dsfont}

\newcommand{\cmark}{\ding{51}}%
\newcommand{\xmark}{\ding{55}}%

\newcommand{\edit}[1]{#1}
\newcommand{\editLS}[1]{#1}

\newcommand{\duoprompt}{DuoPrompt\xspace}
\newcommand{\osl}{$\mathds{1}$SL\xspace}

\AtBeginDocument{%
  \providecommand\BibTeX{{%
    \normalfont B\kern-0.5em{\scshape i\kern-0.25em b}\kern-0.8em\TeX}}}

\copyrightyear{2023} 
\acmYear{2023} 
\setcopyright{acmlicensed}\acmConference[SIGIR '23]{Proceedings of the 46th International ACM SIGIR Conference on Research and Development in Information Retrieval}{July 23--27, 2023}{Taipei, Taiwan}
\acmBooktitle{Proceedings of the 46th International ACM SIGIR Conference on Research and Development in Information Retrieval (SIGIR '23), July 23--27, 2023, Taipei, Taiwan}
\acmPrice{15.00}
\acmDOI{10.1145/3539618.3592032}
\acmISBN{978-1-4503-9408-6/23/07}

\begin{document}

\title[One-Shot Labeling for Automatic Relevance Estimation]{One-Shot Labeling for Automatic Relevance Estimation}

\author{Sean MacAvaney}
\authornote{Listed alphabetically; both authors contributed equally to this research.}
\affiliation{%
  \institution{University of Glasgow}
  \country{United Kingdom}
}
\email{sean.macavaney@glasgow.ac.uk}
\author{Luca Soldaini}
\authornotemark[1]
\affiliation{%
  \institution{Allen Institute for AI}
  \country{United States}
}
\email{lucas@allenai.org}

\renewcommand{\shortauthors}{Sean MacAvaney \& Luca Soldaini}
\begin{abstract}
Dealing with unjudged documents (``holes'') in relevance assessments is a perennial problem when evaluating search systems with offline experiments. Holes can reduce the apparent effectiveness of retrieval systems during evaluation and introduce biases in models trained with incomplete data. In this work, we explore whether large language models can help us fill such holes to improve offline evaluations. We examine an extreme, albeit common, evaluation setting wherein only a single known relevant document per query is available for evaluation. We then explore various approaches for predicting the relevance of unjudged documents with respect to a query and the known relevant document, including nearest neighbor, supervised, and prompting techniques. We find that although the predictions of these One-Shot Labelers (\osl) frequently disagree with human assessments, the labels they produce yield a far more reliable ranking of systems than the single labels do alone. Specifically, the strongest approaches can consistently reach system ranking correlations of over 0.86 with the full rankings over a variety of measures. Meanwhile, the approach substantially \edit{increases the reliability} of $t$-tests due to \edit{filling} holes in relevance assessments, giving researchers more confidence in results they find to be significant. 
\editLS{Alongside this work, we release an easy-to-use software package to enable the use of \osl for evaluation of other ad-hoc collections or systems.}

\vspace{0.6em}
\hspace{3.8em}\includegraphics[width=1.25em,height=1.25em]{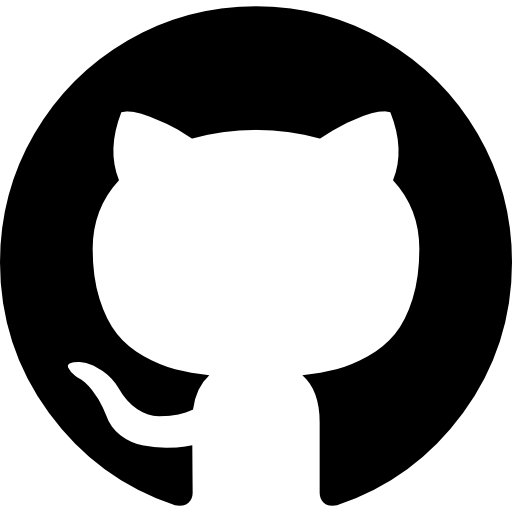}\hspace{.3em}
\parbox[c]{\columnwidth}
{
    \vspace{-.55em}
    \href{https://github.com/seanmacavaney/autoqrels}{\nolinkurl{github.com/seanmacavaney/autoqrels}}
}
\vspace{-1.2em}

\end{abstract}



\maketitle

\section{Introduction}

Cranfield-style test collections~\cite{Cleverdon1997TheCT}---consisting of a document corpus, a set of queries, and manually-assessed relevance judgements between queries and documents---are the bedrock of most information retrieval research. 
Since it is often prohibitively expensive to assess every document for every query (for a non-trivially sized corpus), a subset of documents are chosen for each query for assessment. 
This is typically accomplished through a pooling process, wherein the top results from one or more systems are selected for assessment. 
Decisions about pooling (\textit{e.g.}, how many systems, how many results per system, etc.) have a direct effect on the cost of assessment in terms of human labor. 
On one extreme, some collections only pool from the very top results (\textit{e.g.}, top 10) of a single system. 
This makes the cost of assessing a single query low, but results in very incomplete assessments~\cite{arabzadeh-2022-shallow} and survivorship bias~\cite{gupta:sigir2022-survivor}. At the other extreme, some collections assess the tens to hundreds of documents from dozens of systems. 
This comes at a much higher cost, but results in test collections that are more complete, improving test collection reusability since documents returned from new systems are more likely to already be assessed~\cite{voorhees:2021}. 
Given an identical assessment budget, arguments have been made in favor of both techniques. 
While reusability of deeper pools is an appealing advantage, shallower assessment means that more queries can be assessed, improving coverage of various information needs and making statistical significance easier~\cite{boytsov:2013}. 
Shallow assessments can also be more easily collected, since they do not require a diverse set of input systems, and can be inferred from query logs (e.g.,~\cite{rekabsaz:2021}), making them popular in industry and in large-scale collections like MS MARCO~\cite{Campos2016MSMA}. 
Further, even with deep assessments, ``holes'' (missing assessments) still exist, especially for large corpora, where many documents can satisfy a query's information need~\cite{voorhees:2022}.

In this work, we explore the extent to which we can fill holes in relevance assessments using models over query and document text. 
To reign in the scope, we focus on a setting where only a single relevant document per query, sampled from the top results of a baseline system, is known. 
This setup is akin to the MS MARCO dataset, and is often used in industry. 
We explore several ``one-shot'' approaches\footnote{Zero-shot methods are simply relevance models, which could be used as ranking models directly. 
Since these would clearly favor systems that use them for ranking, we only consider approaches that make use of a known relevant document.}~\editLS{(henceforth \textit{one-shot labelers}, or \osl)} for predicting the relevance of holes, including a $k$ nearest neighbor search of the known relevant document, a supervised relative relevance scorer~(DuoT5~\cite{Pradeep2021TheED}), and an instruction-tuned model~(FlanT5~\cite{Chung2022ScalingIL}). 
Through experiments on the TREC DL 2019, 2020, and 2021 datasets, we find that \osl approaches are not yet able to sufficiently identify all relevant documents (best F1 score: 0.63). 
Therefore, \osl is likely inadequate for estimating system recall. 
However, we find that when incorporating the estimated relevance scores into recall-agnostic C/W/L measures~\cite{moffat:2017}, we are able to achieve high correlation with the ranking of systems from full TREC assessments (consistently $\ge0.86$ correlation, often $\ge0.98$). 
Further, we find that using these approaches yields \edit{more reliable} statistical tests when comparing system performance. 
These results suggest that automatic \osl techniques could potentially replace expensive deep assessments when evaluating with precision-oriented measures.

\section{Related Work}

The Cranfield evaluation paradigm~\cite{Cleverdon1997TheCT} was introduced as a more cost-effective alternative to a user study: 
instead of asking a population of users to interact with IR systems, use expert assessors to judge whether documents retrieved by IR systems are relevant for a set of predefined topics, each representing an hypothetical information need.
Cranfield evaluations rely on a series of assumptions~\cite{Voorhees2001ThePO} to justify this simplification: (\textit{i}) relevance can be approximated as similarity between documents and topics, (\textit{ii}) the topics are a representative approximation of user information need, and, most pertinent to our work, (\textit{iii}) all relevant documents for a query are known. 
The latter requirement is often violated, as it is unfeasible to judge every document in a collection containing  up to billions of items~\cite{Zobel1998HowRA,Voorhees2001ThePO}. 
Instead, pooling techniques are often used to create a test collection: all systems part of a shared task contribute\footnote{Many strategies can be employed to sample documents from the output of a IR system in order to build a pool.} to a pool of documents~\cite{SparckJones1975ReportOT};
items in the pool get judged by annotators; any document not in the pool is assumed to be irrelevant. 

\textbf{Metrics for Partially Assessed Collections.}
In response to the inevitable violation of completeness assumption in pooled evaluation, researchers have proposed metrics that take this limitation into account.
Early on, \citet{Buckley2004RetrievalEW}~measured the effect of this violation in common metrics, such as mean Average Precision, Precision@k retrieved results, and R-Precision; 
further, they proposed Bpref, a metric that exhibits fewer errors in ranking systems that contributed to the evaluation pool. 
Follow up work proposed rank-biased precision (RBP)~\cite{Moffat2007StrategicSC, Moffat2008RankbiasedPF} and inferred precision~\cite{Yilmaz2006EstimatingAP,Sakai2007AlternativesTB};
the latter has been argued to be more effective than Bpref and RBP thanks to its superior discriminative power~\cite{Sakai2008OnIR}. 
\edit{\citet{DBLP:conf/ecir/FrobeGPH23}~proposed using bootstrapping to overcome holes.}

Despite these efforts, the community has largely not adopted these measures;\footnote{As a few examples, none of the TREC 2021 tracks reported measures that were robust to unjudged documents as official evaluation measures, nor did any of papers at SIGIR 2022 ``Search and Ranking'' session.} instead, measures that simply treat unjudged documents as irrelevant---such as nDCG, MAP, and P@k---remain in common use. Meanwhile, it has also become increasingly popular to simply report the proportion of documents in the top results that have relevance assessments (Judged@k, or conversely Hole@k) alongside these measures (e.g.,~\cite{MacAvaney2019TeachingAN,Thakur2021BEIRAH}).
Recognizing the preference for traditional metrics that rely on full judgments, our \osl{s} can effectively fill holes in the evaluation pool, thus mitigating violations of the completeness assumption.

\textbf{Automated Assessment of Documents.}
Given the limitations arising from incomplete judgments, automated tools have been proposed to  assist or automate the evaluation of retrieval systems. Some researchers have focused on completely automating IR evaluation~\cite{Soboroff2001RankingRS,Wu2003MethodsFR,wu2001datafusion,NurayTuran2006AutomaticRO,Spoerri2007UsingTS};
however, these works have been generally criticized as evaluating retrieval systems by ``popularity as opposed to performance''~\cite{Aslam2003OnTE}. 
Further, recent studies of fully-automated relevance estimators found that careful selection of topics~\cite{Hauff2010ACF} and ensembling of estimators~\cite{Roitero2020EffectivenessEW} are necessary to best employ these otherwise unstable techniques.

Rather that completely relying on models, our work instead focuses on how to best exploit a very small number of relevance judgments provided by human assessors to label an otherwise unusable collection. 
Previous work in this area is built upon the cluster hypothesis~\cite{DBLP:journals/ipm/JardineR71}: 
that is, it assumes that documents that are similar to relevant documents must also be relevant.
For example, \citet{Carterette2007SemiautomaticEO}~used \textit{tfidf}-weighted vector space similarity to estimate the probability of a unjudged document.
\citet{Bttcher2007ReliableIR}~experimented with predicting relevance of documents using models that estimate unigram distribution over relevant and non-relevant documents. \edit{Another direction involves human-written criteria, which can be automatically assessed by QA systems~\cite{DBLP:conf/desires/SanderD21}.} \citet{Hui2016ClusterHI} evaluates different strategies to measure document similarity for MaxRep, concluding that bag of words and paragraph embedding approaches to be most suitable.
In this manuscript, we extend their approach by using modern contextualized embedders, and compare it with our proposed approaches.

Outside adhoc retrieval, others have proposed methods for completing partially annotated pools for temporal summarization~\cite{McCreadie2018AutomaticGT}, 
novelty-oriented retrieval~\cite{Hui2017DealingWI}, and question answering~\cite{Vu2021AVAAA} tasks.

\begin{table}[t]
\centering
\caption{Comparison of proposed gain estimators. \xmark* indicates a system that was not directly supervised on the data, but may seen some during instruction tuning.}
\label{tab:overview}
\begin{tabular}{lccccc}
\toprule
& \multicolumn{3}{c}{Considers} & MS MARCO & \\
\cmidrule(lr){2-4}
$G$ & $q$ & $d^+$ & $d^?$ & Supervised & \# Params \\
\midrule
MaxRep-BM25  & \xmark & \cmark & \cmark & \xmark  & n/a \\
MaxRep-TCT   & \xmark & \cmark & \cmark & \cmark  & 110M \\
DuoT5        & \cmark & \cmark & \cmark & \cmark  & 3B \\
\duoprompt   & \cmark & \cmark & \cmark & \xmark* & 3B \\
\bottomrule
\end{tabular}
\end{table}

\section{Automatic One-Shot Labeling (\osl)}
\label{sec:one_shot_labeling}

Consider the case where only a single relevant document $d^+$ per query $q$ is known. Many common evaluation measures (\textit{e.g.}, those in the C/W/L framework~\cite{moffat:2017}) begin by mapping a system ranking $[d_1, d_2, ..., d_n]$ to gain values $[g_1, g_2, ..., g_n]$. It is common practice to treat documents that are not assessed as having no gain, \textit{i.e.},
\begin{equation}\small
g_i =
\begin{cases}
    1 & \text{if } d_i = d^+\\
    0 & \text{otherwise}
\end{cases}
\end{equation}
In this work, we explore techniques for estimating the gain for an unknown document in the ranking $d^?$ (i.e., ``holes'') with respect to $d^+$ and $q$ using a relevance estimator function $G(q,d^+,d^?)\in[0,1]$:\footnote{Without loss of generality, we assume the outputs are in the range of 0 to 1; in cases where they are not, the outputs could simply be normalized using various approaches.}
\begin{equation}\small
g_i =
\begin{cases}
    1 & \text{if } d_i = d^+\\
    G(q, d^+, d_i) & \text{otherwise}
\end{cases}
\end{equation}
These gains can then be used by various evaluation measures. We refer to $G$ as a One-Shot Labeler (\osl), given that it provides labels for a query based on a single known relevance label. We now describe several \osl implementations, summarized in Table~\ref{tab:overview}.

\textbf{MaxRep.} We adapt MaxRep by~\citet{Hui2015SelectiveLA} to a one-shot setting: 
given $d^+$, we retrieve the $k$ nearest neighboring documents and also treat them as relevant, with a linearly degrading gain (\textit{i.e.}, the $i^\text{th}$ nearest neighbor receives a gain of $(k-i)/k$). Note that this approach does not use the query directly, except insofar as that $d^+$ is known to be relevant to $q$. In line with prior work~\cite{Hui2016ClusterHI}, we explore both lexical similarity and semantic similarity. For lexical similarity, we use the top $k=128$ BM25 results for the given document, and for semantic similarity, we use the top $k=128$ nearest neighbors over TCT-ColBERT~\cite{lin-etal-2021-batch} embeddings\footnote{\texttt{castorini/tct\_colbert-v2-hnp-msmarco-r2}} --- both of which have been recently shown as an effective signal to identify additional relevant documents when re-ranking~\cite{macavaney:cikm2022-adaptive}.

\textbf{DuoT5.} The DuoT5 model~\cite{Pradeep2021TheED} is a sequence-to-sequence model trained to provide a relative relevance signal between two documents, with respect to a query. It was proposed for use as a final stage re-ranker, where it can help refine the top documents in a ranked list by comparing them with one another. We recognize that it may also be suitable estimating the relevance in a one-shot setting. Specifically, by presenting the model with $q$ and $d^+$, the model estimates the probability that $d^?$ is more relevant to $q$ than $d^+$. Note that in contrast with when DuoT5 is used as a re-ranker, this case makes use of a \textit{known} relevant document, and uses the information to predict the relevance of additional documents.

\textbf{\duoprompt.}
Recent advances in instruction-tuned models~\cite{Wei2021FinetunedLM,Sanh2021MultitaskPT,Wang2022SuperNaturalInstructionsGV} have lead to the rise of models that can be prompted ``in context'' to complete a task. 
Practically, this consists of prefixing each sample one wishes to perform inference on with instructions describing the task. 
In our case, we prepend query and passage with a short text prompting the model to estimating relevance of passages for the given query.\footnote{Prompt: \texttt{Determine if passage B is as relevant as passage A. Passage A: <d+> Passage B: <d?> Query: <q> Is passage B as relevant as passage A?}} Akin to DuoT5, \duoprompt prompts the model to assess whether $d^{?}$ is as relevant as  $d^{+}$ for $q$. 
We teacher-force the model to generate either ``yes'' or ``no''; the probability values obtained by computing the softmax over the logits of these two tokens is used as the relevance score.\footnote{Inspired by its success in prior work~\cite{Min2022RethinkingTR,Zhao2021CalibrateBU,Liu2021WhatMG}, we also experimented with adding examples of the tasks to the prompt, a practice known as in-context learning.
However, we found it to be worse than not including any labeled samples. 
This might be due to the large variance in passages, but we leave the problem of exploring in-context learning to future work.} 
We leverage the pre-trained Flan-T5~\cite{Chung2022ScalingIL}, a variant of T5~\cite{Raffel2019ExploringTL} that has been fine-tuned on a large set of instruction datasets.\footnote{We note that the instruction data includes \edit{500} samples from the MS MARCO QnA \edit{train} dataset, resulting in a model that is supervised on a small amount of data from our target domain. \edit{However, the instruction prompts do not make use of the relevance labels; only a set of candidates, the queries and assessor-written answers.}}

\section{Experiments and Results}

\begin{figure}
\centering
\includegraphics[scale=0.68]{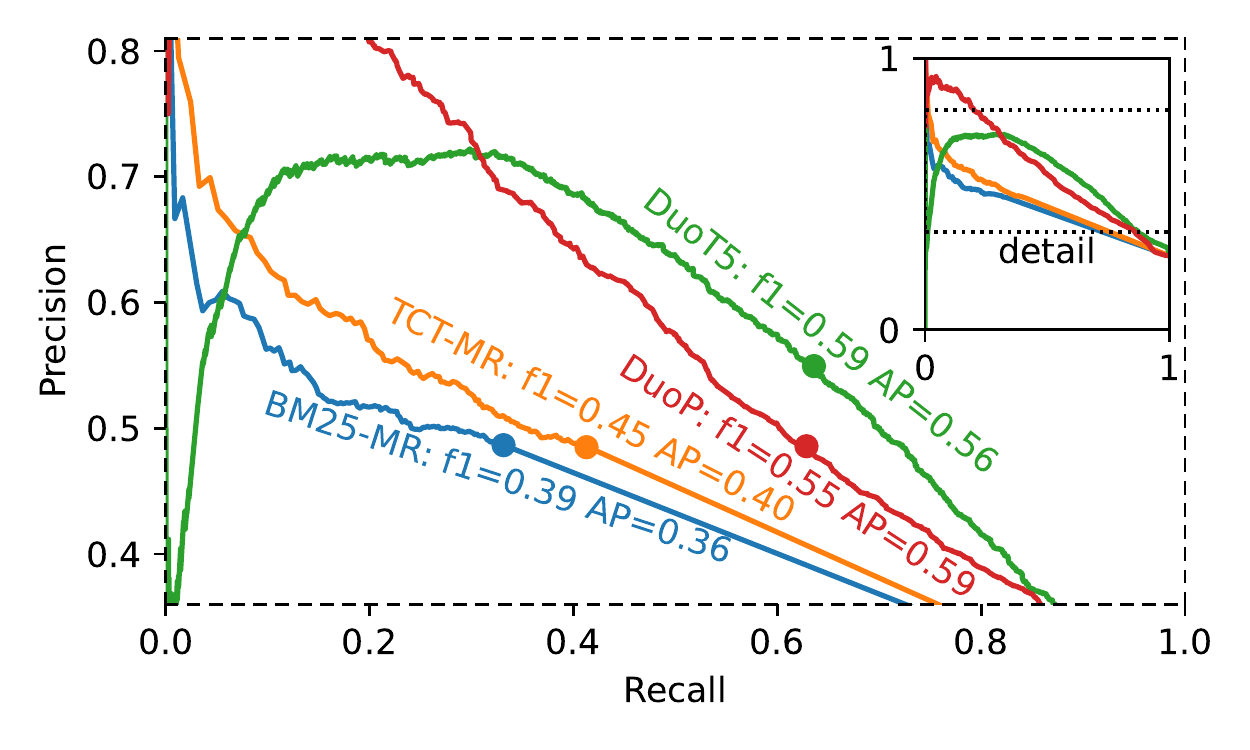}
\vspace{-1em}
\caption{Precision-recall curve of one-shot labelers (\osl{s}) on the \textit{qrels} from the 2019 TREC Deep Learning Track.}
\label{fig:pr_curve}
\end{figure}

\input{tab_main}

In this section, we establish two research questions and evaluate how one-shot modelers perform in these two settings. 
First, \textbf{RQ1} asks: \textit{can \osl{s} be used to directly label unjudged documents in a shared task pool?}
While appealing, we show that proposed techniques cannot reliably provide binary relevance labels.
Then, \textbf{RQ2} asks (despite the limitations described above): \textit{can \osl{s} be used to rank IR systems a shared task?}
Crucially, we find that \osl{s} are very accurate relative performance estimators across three adhoc passage retrieval benchmarks and three metrics. 
To answer both questions, we use a simulated shallow document pool, based on the task's official BM25 baseline.\footnote{\texttt{bm25base\_p} for DL 2019, \texttt{p\_bm25} for DL 2020 and 2021} We choose the first relevant document from the run, and treat it as the single $d^+$ for the query.\footnote{On average, this requires 6 passages to be ``examined'' per query.}

\subsection{RQ1: \osl{s} as Relevance Estimators}

\hspace{1.15em}\textbf{Setup.}
\edit{In order to establish whether any of the methods described in Section~\ref{sec:one_shot_labeling} can function as automatic binary relevance assessors, we measure their efficacy at inferring relevance labels for unknown documents. Given the simulated shallow pools, we predict the binary labels of \textit{qrels} from the TREC DL 2019 track~\cite{craswell-2019-overview}.}

\textbf{Results.} \edit{We report the precision-recall curves of these predictions in~Figure~\ref{fig:pr_curve}.}
We note clear differences among the methods introduced in Section~\ref{sec:one_shot_labeling}.  
First, MaxRep methods achieve notably worse F1 and Average Precision (AP) than other methods.
This is likely due to MaxRep not considering query text when estimating relevance, instead only relying on \edit{intra-document} similarity. On the other hand, \duoprompt and DuoT5 are moderately effective at the tasks, achieving an F1-score of at most $0.55$ and $0.59$ (AP: $0.59$ and $0.56$) respectively. 
However, these results are far from meeting necessary accuracy to use \duoprompt or DuoT5 as fully automatic labelers, as they are unable to reliably identify the long tail of relevant documents. 
\edit{Further, we note that their PR curves are very different, making difficult to tune a reliable operating point for the binary labels.}
This \edit{observation} confirms that the difficulties noted by~\citet{Hauff2010ACF} and~\citet{Roitero2020EffectivenessEW} in tuning zero-shot labelers applies to \edit{our} \osl{}s.

\subsection{RQ2: \edit{\osl{s} for System Evaluation}}

\hspace{1.15em}\textbf{Setup.}
In order for \osl{s} to be a viable alternative to highly skilled assessors, they should lead to similar evaluation outcomes. 
More specifically, given a set of systems submitting runs to a shared track, results using \osl{s} should result in a ranking of participating systems that is comparable to that of human labels \edit{and for similar outcomes from statistical significance tests}.

To study if that is the case, we calculate various evaluation measures for the submissions to the TREC 2019, 2020, and 2021 Deep Learning tracks~\cite{craswell-2019-overview,craswell-2020-overview,craswell-2021-overview} \edit{(37, 58, and 58 runs, respectively)} using our simulated shallow pools. We treat holes as either non-relevant (0-gain, the baseline), or by replacing the relevance gains using \osl scores. Since we found that \osl{s} are not yet capable of providing a reliable measure of recall, we instead focus on three recall-agnostic evaluation measures from the C/W/L framework: SDCG@10, P@10, and RBP(p=0.8) \edit{using \texttt{ir-measures}~\cite{DBLP:conf/ecir/MacAvaneyMO22a}}.\footnote{\edit{nDCG@10 and P@10 are official TREC DL task metrics.} SDCG@10 is similar to nDCG@10, but assumes there are 10 fully-relevant documents available (and is therefore unaffected by the introduction of new relevant documents \edit{to the qrels}).} We use variants of P@10 \edit{(\textit{weighted-precision})} and RBP that make use of partial gains\edit{~\cite{moffat:2017}}.

\textbf{Results.}
We report our main findings in Table~\ref{tab:main}.
Here the first row in each section corresponds to estimating system ranks using only the one labeled document from BM25; subsequent rows show how \osl{s} perform. Overall, we observe that \duoprompt usually leads to more accurate estimation across all years and metrics:
it achieves a correlation of $0.87$ to $0.92$, and a rank correlation of $0.97$ to $0.98$. 
More importantly, $t$-tests conducted using the method are far more reliable than other methods\edit{; it rarely gives false positive results (as compared to the full \textit{qrels}), and infrequently yields false negatives}. DuoT5 also often achieves high correlations with the full qrels, but \edit{results in more false positive $t$-tests}. On the other hand, MaxRep~\cite{Hui2015SelectiveLA} does not offer reliable estimation of labels, leading to inferior correlations when compared to DuoT5 and \duoprompt. 
MaxRep-BM25 is particularly unreliable, with all its correlation~$<0.5$; 
MaxRep-TCT, while more accurate than MaxRep-TCT, still suffers from \edit{higher false positive rates in most cases than \duoprompt}.
Finally, while unsurprising, we note that, ordering of systems induced by using only one positive labels (``not relevant'' in  Table~\ref{tab:main}) is a very inaccurate approach, leading to very low or even negative correlations and a high proportion of false positive tests. \edit{Clearly, in this case, the single system that contributed to the \textit{qrels} is unfairly favored above the other systems, leading to artificially higher evaluation measures than it would under deeper pools.}

\section{Conclusions and Recommendations}

In this work, we presented a new approach to replace deep manual relevance assessments with prediction from automatic one-shot labelers (\osl{s}). 
These models rely on one relevant document per topic to evaluate a pool of documents retrieved by IR systems. 
This approach can reliably predict ordering of systems across multiple shared tasks and metrics.
One obvious question is: \textbf{Can shared tasks organizers largely do away with human labelers?}\footnote{\edit{We refer the readers to \citet{Faggioli2022Perspectives}, a contemporaneous article that explores this question in greater detail. Here, we answer this question with respect to our study.}}

First, we emphatically stress that whether to completely replace human assessors \textbf{is not} the right question to ask; 
rather, accurate automatic labelers allow re-distributing existing annotation budgets in ways that unlock new evaluation practices. 
First, our approach requires at least one high-quality relevant document.
More importantly, \osl{s} can be used in conjunction with human annotators to obtain a pool that is both wide and deep: human annotators can be tasked to label a few documents for each topic in a large set of queries, with automatic labelers inferring judgements on the reminder of the pool.
This would lead easier measurement of significant improvements~\cite{boytsov:2013}, and the ability to study more complex information needs.
Moreover, automatic labelers could increase value of existing test collections: \edit{an estimation of the relevance of}
any previously unjudged document retrieved by new systems could be \edit{used instead of treating them as irrelevant}.

We note that the scope of this study has limitations that would need to be explored before this approach be used in practice. 
We only explore passage retrieval; providing relative relevance estimates over entire documents is likely a far more challenging task, especially with context window limitations of neural language models. 
Further, we only explore situations where a single relevant document is known; rather, we often know a set of relevant documents (of potentially differing relevance grades). Techniques for using multiple labels (e.g., aggregating scores from multiple relevant documents) need to be explored. \edit{We also dodged the issue of measures that rely on recall by using the utility-based C/W/L measures. Although such measures are practical in our setting, measures that depend on system recall (such as nDCG, MAP, etc.) are commonly used in evaluations.}
Finally, more work is needed to explore potential biases these models introduce, especially when the systems share a base language model.

\bibliographystyle{ACM-Reference-Format}
\balance
\bibliography{sample-base}

\end{document}

%% file: tab_main.tex
\begin{table*}
\centering
\small
\renewcommand{\arraystretch}{0.9}
\caption{\osl{s} as system evaluators. Correlations of the TREC submissions are reported in terms of Kendall's $\tau$, Spearman's $\rho$, Rank Biased Overlap ($p=0.9$), and the $t$-test ($p<0.05$ w/ Bonferroni) false positive/negative rates of the top identified system.}
\label{tab:main}
\vspace{-1em}
\scalebox{0.93}{
\begin{tabular}{ll|rrrrr|rrrrr|rrrrr}
\toprule
\multicolumn{2}{c}{} & \multicolumn{5}{c}{\tt msmarco-passage/trec-dl-2019}& \multicolumn{5}{c}{\tt msmarco-passage/trec-dl-2020}& \multicolumn{5}{c}{\tt msmarco-passage-v2/trec-dl-2021}\\
\cmidrule(lr){3-7}\cmidrule(lr){8-12}\cmidrule(lr){13-17}
Measure & \multicolumn{1}{l}{Holes} & $\tau$ & $\rho$ & RBO & \multicolumn{1}{r}{$t$-FNR} & \multicolumn{1}{r}{$t$-FPR} & $\tau$ & $\rho$ & RBO & \multicolumn{1}{r}{$t$-FNR} & \multicolumn{1}{r}{$t$-FPR} & $\tau$ & $\rho$ & RBO & \multicolumn{1}{r}{$t$-FNR} & \multicolumn{1}{r}{$t$-FPR} \\
\midrule

\textbf{SDCG@10} & Non-relevant & -0.204 & -0.248 & 0.420 &\bf  0.000 & 0.857 & 0.419 & 0.498 & 0.731 &\bf  0.000 & 0.510 & 0.402 & 0.531 & 0.564 & 0.125 & 0.714 \\                                                     
 & MaxRep-BM25 & 0.240 & 0.320 & 0.690 &\bf  0.000 & 0.229 & 0.468 & 0.597 & 0.710 &\bf  0.000 & 0.059 & 0.455 & 0.576 & 0.609 & 0.200 & 0.489 \\                                                                        
 & MaxRep-TCT & 0.829 & 0.958 & 0.818 & 0.208 & 0.083 & 0.793 & 0.933 & 0.920 & 0.263 & 0.053 & 0.578 & 0.761 & 0.817 & 0.222 & 0.300 \\                                                                         
 & DuoT5 & 0.889 & 0.972 & 0.812 &\bf  0.000 & 0.417 & 0.837 & 0.944 &\bf  0.939 &\bf  0.000 & 0.895 & 0.859 & 0.963 & 0.880 &\bf  0.000 & 0.571 \\                                                                              
 & DuoPrompt &\bf  0.904 &\bf  0.980 &\bf  0.830 & 0.160 &\bf  0.000 &\bf  0.909 &\bf  0.986 & 0.863 & 0.184 &\bf  0.000 &\bf  0.910 &\bf  0.983 &\bf  0.925 & 0.040 &\bf  0.143 \\                                                                          
\midrule                                                                                                                                                                                                                                       
\textbf{P@10} & Non-relevant & -0.033 & -0.003 & 0.511 &\bf  0.000 & 0.571 & 0.429 & 0.526 & 0.695 & 0.143 & 0.120 & 0.387 & 0.520 & 0.528 & 0.182 & 0.739 \\                                                        
 & MaxRep-BM25 & 0.362 & 0.452 & 0.722 &\bf  0.000 & 0.314 & 0.442 & 0.565 & 0.660 &\bf  0.000 & 0.118 & 0.425 & 0.579 & 0.571 & 0.167 & 0.467 \\                                                                        
 & MaxRep-TCT & 0.870 & 0.971 &\bf  0.874 & 0.208 &\bf  0.083 & 0.792 & 0.928 & 0.769 & 0.214 & 0.067 & 0.579 & 0.767 & 0.781 & 0.154 & 0.290 \\                                                                         
 & DuoT5 &\bf  0.891 & 0.974 & 0.811 &\bf  0.000 & 0.333 & 0.868 & 0.970 & 0.781 &\bf  0.000 & 0.867 & 0.858 & 0.962 &\bf  0.916 &\bf  0.000 & 0.700 \\                                                                              
 & DuoPrompt &\bf  0.891 &\bf  0.981 & 0.872 &\bf  0.000 & 0.250 &\bf  0.907 &\bf  0.986 &\bf  0.817 & 0.143 &\bf  0.000 &\bf  0.868 &\bf  0.972 & 0.903 &\bf  0.000 &\bf  0.200 \\                                                                          
\midrule                                                                                                                                                                                                                                       
\textbf{RBP(p=0.8)} & Non-relevant & -0.177 & -0.220 & 0.445 &\bf  0.000 & 0.857 & 0.437 & 0.510 & 0.840 &\bf  0.000 & 0.353 & 0.387 & 0.516 & 0.535 & 0.167 & 0.706 \\                                                  
 & MaxRep-BM25 & 0.246 & 0.332 & 0.713 &\bf  0.000 & 0.200 & 0.452 & 0.583 & 0.715 &\bf  0.000 & 0.059 & 0.446 & 0.562 & 0.601 & 0.182 & 0.565 \\                                                                        
 & MaxRep-TCT & 0.853 & 0.963 & 0.829 & 0.167 & 0.083 & 0.791 & 0.930 & 0.917 & 0.333 &\bf  0.000 & 0.569 & 0.761 & 0.766 & 0.259 & 0.333 \\                                                                         
 & DuoT5 & 0.892 & 0.975 & 0.854 &\bf  0.000 & 0.667 & 0.806 & 0.929 & 0.939 &\bf  0.000 & 0.889 & 0.863 & 0.970 & 0.882 &\bf  0.000 & 0.625 \\                                                                              
 & DuoPrompt &\bf  0.919 &\bf  0.986 &\bf  0.889 & 0.040 &\bf  0.000 &\bf  0.889 &\bf  0.980 &\bf  0.964 & 0.179 &\bf  0.000 &\bf  0.890 &\bf  0.980 &\bf  0.897 &\bf  0.000 &\bf  0.250 \\

\bottomrule
\end{tabular}
}
\end{table*}